\documentstyle[sprocl]{article}
\input{psfig}
\def\be{\begin{equation}}
\def\ee{\end{equation}}
\def\bea{\begin{eqnarray}}
\def\eea{\end{eqnarray}}

\def\bra{\langle}
\def\ket{\rangle}
\def\a{\alpha}
\def\b{\beta}
\def\g{\gamma}

\def\p{\pi}

\def\l{\lambda}
\def\m{\mu}
\def\n{\nu}
\def\G{\Gamma}

\def\to{\rightarrow}
\begin{document}
\thispagestyle{empty}
\begin{flushright}
ITP-SB-97-25, DESY 97-071, April 1997\\
\end{flushright}
\vspace*{0.4cm}
\title{Next-To-Leading Logarithmic Results 
in $B \to X_s \gamma$
\footnote{ Work supported in part by Schweizerischer Nationalfonds,
based on a talk given by T.H. at the Symposium 
{\it Flavor Changing Neutral Currents - Present and Future Studies},
February 19-21, Santa Monica, California, to appear in the Proceedings }}
\author{ CHRISTOPH GREUB }
\address{Deutsches Elektronen-Synchroton DESY,
\\ 22603 Hamburg, Germany}
\author{ TOBIAS HURTH }
\address{Institute for Theoretical Physics, SUNY at Stony Brook,\\
Stony Brook, New York 11794-3840, USA}
\maketitle
\abstracts{
We give a brief review of the next-to-leading logarithmic
results in $B \to X_s \gamma$.
Combining the results of different groups,
a practically complete next-to-
leading-logarithmic  prediction of the inclusive decay rate was 
recently presented.
The theoretical uncertainty in the decay rate is now less than
half of the error in the previously leading-logarithmic result.
Therefore, the inclusive $B \to X_s \gamma$ mode will 
provide important tests 
of the SM and its extensions when more precise experimental data is
available.}
Rare B meson decays provide an alternative approach in the search for
new physics: The $B \to X_s \gamma$ decay in particular
does not arise at the tree level in the standard model (SM) but is  induced 
by one-loop W-exchange diagrams. Therefore nonstandard contributions 
(charged scalar exchanges, SUSY one-loop diagrams etc.) are not suppressed 
by an extra factor $\alpha/4\pi$ relative to the standard model amplitude 
which implies the high sensitivity of this decay for new physics. However, 
even within the SM, the $B \to X_s \gamma$ decay  is also important for constraints 
on the Cabibbo-Kobayashi-Maskawa matrix elements which involve the top-quark. 
For both these reasons, precise experimental and theoretical work
on these decays is required. 

The experimental status of this decay can be summarized as follows:
Following the first observation of the exclusive $B \to K^* \gamma$
mode \cite{Cleo1}, the first evidence for a penguin decay ever, the CLEO collaboration
measured the inclusive  $B \to X_s \gamma$ branching ratio to be 
$(2.32 \pm 0.57 \pm 0.35) \times 10^{-4}$, 
where the first error is statistical and the second
is systematic \cite{Cleo2}. In fact, there are two separate CLEO analyses. The first one
measures the inclusive photon spectrum from B-decay near the end point.
The second technique constructs the inclusive rate by summing up the possible
exclusive final states. The branching ratio stated above is the average 
of the two measurements, taking into account the correlation between
the two techniques.
In the upcoming years much more precise measurements are expected
from the upgraded CLEO detector, as well as from the B-factories
presently under construction at SLAC and KEK. 
In view of the expected high luminosity of the B-factories, experimental
accuracy below $10\%$ seems to be possible. 

The inclusive  $B \to X_s \gamma$ mode in contrast to exclusive decay modes is 
theoretically clean in the sense that no specific model is needed
to describe the final hadronic state. Indeed, heavy quark effective theory tells us that
the decay width $\G(B \to X_s \gamma)$ 
is well approximated by the partonic decay rate
$\G(b\to X_s \gamma)$ which can be
analyzed in renormalization group improved perturbation theory. 
The class of 
non-perturbative effects which scales like $1/m_b^2$
is expected to be well below $10\%$ \cite{Falk}. This numerical
statement is supposed to hold also 
for the recently discovered non-perturbative contributions
which scale like $1/m_c^2$ \cite{Wise}. Therefore, we focus 
on the dominant partonic decay rate in the following.

It is well-known that the  QCD corrections enhance the 
partonic decay rate $ \Gamma(b \to s \g)$  by more than a factor of two. 
These QCD effects can be attributed to logarithms of the form 
$\alpha_s^n(m_b) \, \log^m(m_b/M)$,
where $M=m_t$ or $M=m_W$ and $m \le n$ (with $n=0,1,2,...$).
In order to get a reasonable result at all, one has  to resum at least
the leading-log (LL) series ($m=n$).  
Working to next-to-leading-log (NLL) precision means that one is also resumming all the
terms of the form $\a_s(m_b) \, \left(\a_s^n(m_b) \, \ln^n (m_b/M)\right)$.

An appropriate framework to achieve the necessary resummations 
is an  effective 
low-energy theory, obtained by integrating out the
heavy particles which in the SM are the top quark and the $W$-boson. 
The effective Hamiltonian relevant for $b \to s \gamma$ and
$b \to s g$ in the SM and most of its extensions reads 
\begin{equation}
\label{heff}
H_{eff}(b \to s \gamma)
       = - \frac{4 G_{F}}{\sqrt{2}} \, \lambda_{t} \, \sum_{i=1}^{8}
C_{i}(\mu) \, O_i(\mu) \quad ,
\end{equation}
where $O_i(\m)$ are the relevant operators,
$C_{i}(\mu)$ are the corresponding Wilson coefficients,
which contain the complete top- and W- mass dependence,
and $\lambda_t=V_{tb}V_{ts}^*$ with $V_{ij}$ being the
CKM matrix elements \footnote{The CKM dependence globally factorizes,
because we work in the approximation $\l_u=0$.}.
Neglecting operators with dimension $>6$ which are suppressed 
by higher powers of $1/m_{W/t}$ and using the equations
of motion for the operators, one arrives at the following basis 
of dimension 6 operators \cite{Grinstein90}
\bea
\label{operators}
O_1 &=& \left( \bar{c}_{L \b} \g^\m b_{L \a} \right) \,
        \left( \bar{s}_{L \a} \g_\m c_{L \b} \right)\,, \nonumber \\
O_2 &=& \left( \bar{c}_{L \a} \g^\m b_{L \a} \right) \,
        \left( \bar{s}_{L \b} \g_\m c_{L \b} \right) \,,\nonumber \\
O_7 &=& (e/16\p^{2}) \, \bar{s}_{\a} \, \sigma^{\m \n}
      \, (m_{b}(\mu)  R + m_{s}(\mu)  L) \, b_{\a} \ F_{\m \n} \,,
        \nonumber \\
O_8 &=& (g_s/16\p^{2}) \, \bar{s}_{\a} \, \sigma^{\m \n}
      \, (m_{b}(\mu)  R + m_{s}(\mu)  L) \, (\l^A_{\a \b}/2) \,b_{\b}
      \ G^A_{\m \n} \quad .
\eea
Because the Wilson coefficients of the penguin induced Four-Fermi
operators $O_3,..O_6$ are very small, we do not list them here.
The perturbative QCD corrections for the $b \to s \gamma$ decay rate are 
twofold:\\ 
{\bf 1} $\bullet$ The corrections to the Wilson coefficients $C_i(\mu)$ at the 
scale $\mu \approx m_b$.\\
 {\bf 2} $\bullet$ The corrections to the matrix elements of the operators $O_i$ 
also at the low-energy scale $\mu \approx m_b$.\\
Only the sum of the two 
contributions is renormalization scheme 
independent and in fact, 
from the $\mu$-independence of the effective Hamiltonian,
one can derive a renormalization group equation 
(RGE) for the Wilson 
coefficients $C_i(\mu)$:
\be
\label{RGE}
\mu \frac{d}{d\mu} C_i(\mu) = \gamma_{ji} \, C_j(\mu) \quad ,
\ee  
where the $(8 \times 8)$ matrix $\gamma$ is the anomalous dimension
matrix of the operators $O_i$.
The standard procedure to calculate the two contributions involves the
following three steps:\\
{\bf ad 1a} $\bullet$  One has to match the full standard model theory
with the effective theory at the scale $\m=\m_W$, where
$\m_W$ denotes a scale of order $m_W$ or $m_t$. At this scale,
the matrix elements of the operators  in the 
effective theory lead to the  same logarithms  as the full theory
calculation. 
Consequently, the Wilson coefficients 
$C_i(\m_W)$ only pick up small QCD corrections,
which can be calculated in fixed-order perturbation theory.
In the LL (NLL) program, the matching has to be worked out to order
$\a_s^0$ ($\a_s^1$).\\
{\bf ad 1b} $\bullet$ Solving the RGE (\ref{RGE}) and using the $C_i(\m_W)$ 
of Step 1a as initial conditions, one performs the   
evolution of these Wilson coefficients from 
$\m=\m_W$ down to $\m = \m_b$, where $\m_b$ is of the order of $m_b$.
As the matrix elements of the operators evaluated at the low scale
$\m_b$ are free of large logarithms, the latter are contained in resummed
form in the Wilson coefficients. For a LL (NLL) calculation, this RGE step
has to be performed using the anomalous dimension matrix $\gamma_{ji}$  up 
to order $\a_s^1$ ($\a_s^2$).\\
{\bf ad 2} $\bullet$ The corrections to the matrix elements 
of the operators $\bra s \g |O_i (\mu)|b \ket$ at the scale  $\mu = \m_b$
have to be calculated to order $\a_s^0$ ($\a_s^1$) in the LL (NLL)
calculation.

Until recently, only the leading logarithmic (LL) perturbative QCD
corrections had been calculated 
\cite{counterterm} systematically. The error in these
calculations is dominated by a large 
renormalization scale dependence at the $\pm 25\%$ level. 
The measurement of the CLEO collaboration \cite{Cleo2} overlaps with 
the estimates based on leading logarithmic calculations
(or with some next-to-leading effects partially included)
and the experimental and 
theoretical errors are comparable (see Figure 1) 
\cite{AG91,Burasno,Shifman}. 
\begin{figure} 
\centerline{
\psfig{figure=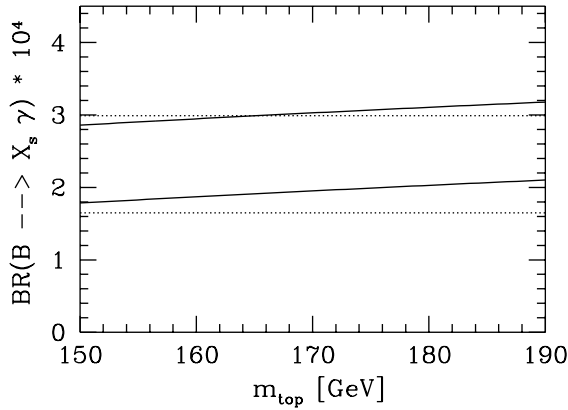,height=2.3in}} 
\caption[]{Branching ratio for $B \to X_s \g$ as a function of $m_t$ based 
on LL calculations.
The upper (lower) solid curve
is for $\mu=m_b/2$ ($\mu = 2 m_b$). The dotted curves show the CLEO
$1-\sigma$ bounds \cite{Cleo2}. The other input parameters
are taken at their central values.}
\end{figure}
However, in view of the expected 
increase in the experimental precision in the near future, 
it is clear that 
a systematic inclusion of the NLL corrections
becomes necessary.  Already the large $\mu$ dependence
of the leading-log result ($\pm 25\%$) indicates the importance
of the NLL series.  This ambitious NLL enterprise was recently completed. 
All three steps (1a,1b,2) to NLL precision involve rather difficult
calculations.   The most difficult part in Step 1a is the 
 two-loop (or  
order $\a_s$) matching of the dipole operators $O_7$ and $O_8$. It involves two-loop
diagrams both in the full and in the effective theory. 
It was worked out by Adel and Yao \cite{Adel} some time ago. 
Using a different method, Greub and Hurth recently presented a 
detailed re-calculation of this step, 
confirming the former result \cite{GGH}.
Step 2 basically consists of Bremsstrahlung corrections and virtual
corrections. While the Bremsstrahlung corrections
(together with some virtual corrections needed to cancel
infrared singularities) were worked
out some time ago by Ali and Greub \cite{AG91} and have
 been confirmed and extended by Pott \cite{Pott}, a  
complete analysis of the virtual corrections (up to the contributions 
of the Four-Fermi operators with very small coefficients) was presented
by Greub, Hurth and Wyler \cite{GHW}. This calculation involves two-
loop diagrams where the full charm dependence has to be taken into account.   
The main result of this analysis
consists in a drastic reduction of the renormalization 
scale uncertainty from about $\pm 25\%$ to about $\pm 5\%$.
Moreover, the central value was shifted outside the $1\sigma$ bound of the
CLEO measurement (see Figure 2).
\begin{figure} 
\centerline{
\psfig{figure=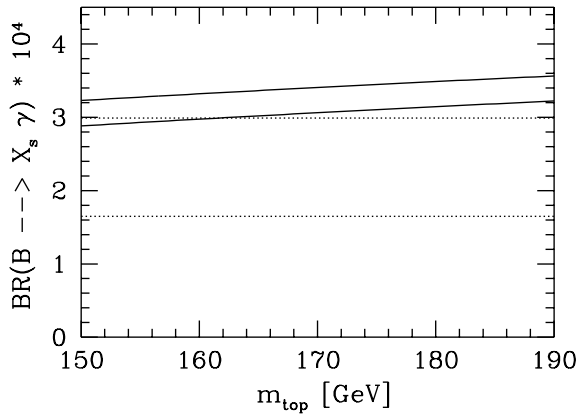,height=2.3in}} 
\caption[]{Branching ratio for $B \to X_s \g$ as a function of $m_t$ based 
on the NLL calculation, not including the NLL corrections to the Wilson
coefficient $C_7$.}
\end{figure}
However, at that time, the essential
coefficient $C_7(\m_b)$ was only known to leading-log precision. 
It was therefore
unclear how much the overall normalization will be changed when
the NLL value for $C_7(\m_b)$ is used. 
Recently, the order $\a_s^2$ anomalous matrix (Step 1b) has been 
completely worked out
by Chetyrkin, Misiak and M\"unz \cite{Mikolaj}. The extraction of some of 
the elements in the $O(\a_s^2)$ anomalous
dimension matrix involves pole parts of three-loop diagrams. 
Using the matching result (Step 1a), these authors obtained 
the next-to-leading correction to the Wilson coefficient $C_7(\m_b)$
which is the only relevant one for the $b \to X_s \gamma$ decay rate.  
Numerically, the LL and the NLL value 
for $C_{7}(\m_b)$ 
are rather similar; the NLL 
corrections to the Wilson coefficient $C_7(\m_b)$ 
lead to a change of the  $b \to X_s \gamma$
decay rate which does not exceed  6\% \cite{Mikolaj}: 
 The new contributions can be split into a part
which is due to the order $\a_s$ corrections to the matching (Step 1a) 
and into
a part stemming from the improved anomalous dimension matrix (Step 1b). 
While individually these two parts are not so small (in the
NDR scheme, which was used in \cite{Mikolaj}), they almost cancel
when combined as illustrated in \cite{Mikolaj}. 
This shows that all the three different pieces are numerically
equally important. However, strictly speaking
the relative importance of different 
NLO-corrections at the scale $\mu = \m_b$, namely 
the order $\alpha_s$ corrections to the matrix elements of 
the operators (Step 2) and the improved
Wilson coefficient $C_7$ (Step 1 a+b), 
is a renormalization-scheme dependent issue; 
so we stress that the discussion  
above was done within  
the naive dimensional regularization scheme (NDR).   

Combining the NLL calculations of all the three steps (1a+b,2), 
the first complete theoretical prediction to NLL  pecision 
for the $b \to X_s + \gamma$ branching ratio 
was presented in \cite{Mikolaj}:
$BR(B \to X_s \g)=(3.28 \pm 0.33) \times 10^{-4}$. The theoretical error 
has two dominant sources:  The 
$\mu$ dependence is reduced to $5\%$ as mentioned above. Another $5\%$ uncertainty stems from the
$m_c/m_b$ dependence.

Summing up, the present NLL-prediction for the $B \to X_s \g$ decay is still
in agreement with the CLEO
measurement at the $2\sigma$-level. The theoretical error is half of the uncertainty in the previous leading logarithmic prediction. Clearly, the inclusive $B \to X_s + \gamma$ 
mode will provide an interesting test of the SM and its extensions
as soon as more precise experimental data are available.

\end{document}